\documentclass[11pt]{iopart}

\usepackage{epsfig,graphicx,latexsym,iopams}
\usepackage{color}
\usepackage{graphicx}  

\def\msun{\,{\rm M_\odot}}
\def\munu{_{\mu \nu}}

\newcommand\be{\begin{equation}}
\newcommand\ee{\end{equation}}
\newcommand{\ba}{\begin{eqnarray}}
\newcommand{\ea}{\end{eqnarray}}

\begin{document}

\title[]{Gravitational wave emission from binary supermassive black holes}

\author{A Sesana$^1$}

\address{$^1$\ Max-Planck-Institut f\"ur Gravitationsphysik, Albert Einstein Institut, Am M\"ulenber 1, 14476 Golm, Germany}

\begin{abstract}
Massive black hole binaries (MBHBs) are unavoidable outcomes of the hierarchical structure formation process, and according to the theory of general relativity are expected to be the loudest gravitational wave (GW) sources in the Universe. In this article I provide a broad overview of MBHBs as GW sources. After reviewing the basics of GW emission from binary systems and of MBHB formation, evolution and dynamics, I describe in some details the connection between binary properties and the emitted gravitational waveform. Direct GW observations will provide an unprecedented wealth of information about the physical nature and the astrophysical properties of these extreme objects, allowing to reconstruct their cosmic history, dynamics and coupling with their dense stellar and gas environment. In this context I describe ongoing and future efforts to make a direct detection with space based interferometry and pulsar timing arrays, highlighting the invaluable scientific payouts of such enterprises.
\end{abstract}


\section{introduction}
\label{intro}
Today, massive black holes (MBHs) are ubiquitous in the nuclei of nearby galaxies ~\cite{mago98}, and we see them shining as quasars along the whole cosmic history up to redshift $z\approx7$ \cite{mortlock11}. In the last decade, MBHs were recognized as fundamental building blocks in hierarchical models of galaxy formation and evolution, but their origin remains largely unknown. In fact, our current knowledge of the MBH population is limited to a small fraction of objects: either those that are active, or those in our neighborhood, where stellar- and gas-dynamical measurements are possible. According to the current paradigm, structure formation proceeds in a hierarchical fashion \cite{wr78}, in which massive galaxies grow by accreting gas through the filaments of the cosmic web and by merging with other galaxies. As a consequence, the MBHs we see in today's galaxies are expected to be the natural end-product of a complex evolutionary path, in which black holes (BHs) seeded in proto-galaxies at high redshift grow through cosmic history via a sequence of MBH-MBH mergers and accretion episodes \cite{kh00,vhm03}. In this framework, a large number of MBH binaries naturally form following the frequent galaxy mergers. 

According to Einstein's theory of General Relativity, accelerating masses cause modifications of the spacetime that propagate at the speed of light, better known as gravitational waves (GWs). However, in Einstein's equations, the matter-metric coupling constant is of the order of $G/c^4$ (where $G$ is the gravitational constant and $c$ is the speed of light), which is of the order $10^{-50}$! As a matter of fact the spacetime is extraordinarily stiff, therefore only massive, compact astrophysical object can produce a sizable strain that would be observable with advanced technology \cite{thorne95}. Two MBHs orbiting each other in a bound binary (MBHB) system carry a huge time varying quadrupole momentum and are therefore expected to be the loudest gravitational wave (GW) sources in the Universe \cite{hughes02}. The frequency spectrum of the emitted radiation covers several order of magnitudes, from the  sub-nano-Hz up to the milli-Hz. The $10^{-4}-10^{-1}$Hz window is going to be probed by spaceborne interferometers like the recently proposed European eLISA \cite{amaro13a,amaro13b,whitepaper13}. At $10^{-9}-10^{-7}$Hz, joint precision timing of several ultrastable millisecond pulsars (i.e. a pulsar timing array, PTA) provides a unique opportunity to get the very first low-frequency detection. The European Pulsar Timing Array (EPTA) \cite{ferdman10}, the Parkes Pulsar Timing Array (PPTA) \cite{man12} and the North American Nanohertz Observatory for Gravitational Waves (NANOGrav) \cite{jenet09}, joining together in the International Pulsar Timing Array (IPTA) \cite{hobbs10}, are constantly improving their sensitivities, getting closer to their ambitious target.  

This focus issue contribution aims at covering all the relevant aspects of MBHBs intended as GW sources, in the spirit of providing a broad overview. Given the extent of the topic, we will just skim through its several facets, providing the appropriate references for in-depth reading. In Section \ref{sec2} we introduce the concept of GWs at a very basic level, defining the relevant astrophysical scales implied by MBHBs. A general overview of MBH formation and evolution (both their masses and spin), together with a brief description of MBHB dynamics is provided in Section \ref{sec3}. We return on GWs in Section \ref{sec4}, where we describe in deeper detail the expected signal from MBHBs and its dependencies on the relevant parameter of the system. There, we also introduce some basics of parameter estimation theory, showing how the rich astrophysical information enclosed in individual signals can be recovered. Section \ref{sec5} is then devoted to the scientific payouts of GW detection both in the milli-Hz and in the nano-Hz regime. We wrap-up in Section \ref{sec6} with some brief conclusion remarks.

\section{Gravitational waves: basics}
\label{sec2}
The existence of GWs was one of the first predictions of Einstein's General Relativity (GR),
since they arise as natural solutions of the linearized Einstein equations in vacuum. Expanding the metric tensor as
\begin{equation}
g\munu=\eta\munu+h\munu,
\end{equation}
where $\eta\munu$ represent the Minkowski flat metric and $\parallel h\munu\parallel\ll\parallel \eta\munu\parallel$, and switching to the appropriate Lorentz gauge, the perturbation $h\munu$ satisfies
\begin{equation}
\Box {h}\munu = -{16\pi G\over c^4}\,T\munu,
\label{einsteinlin}
\end{equation}
where $\Box$ is the d'Alambertian operator and the source term $T\munu$ is the stress-energy tensor. Equation \ref{einsteinlin} represents a set of wave equations and therefore admits wave solutions. These solutions are ripples in the fabric of the spacetime propagating at the speed of light: GWs. GWs are transverse, i.e. they act in a plane perpendicular to the wave propagation, and (at least in GR) have two distinct polarizations, usually referred to as $h_+$ and $h_\times$ (see cartoon in \cite{thorne95} and Section \ref{sec4.2}). By expanding the mass distribution of the source into multipoles, conservation laws enforce GWs coming from the mass monopole and mass dipole to be identically zero, so that the first contribution to GW generation comes from the mass quadrupole moment $Q$. The GW amplitude is therefore proportional to the second time derivative (acceleration) of $Q$. Moreover, energy conservation enforces the amplitude to decay as the inverse of the distance to the source, $D$. A straightforward dimensional analysis shows that the amplitude of a GW is of the order of \cite{hughes03}
\begin{equation}
h=\frac{G}{c^4}\frac{1}{D}\frac{d^2Q}{dt^2}.
\label{eqstrain}
\end{equation}
In order to generate GWs we therefore need accelerating masses with a time varying mass-quadrupole moment. The prefactor $G/c^4$ implies that these waves are {\it tiny}, so that the only detectable effect is produced by massive compact astrophysical objects. 

We now specialize to the case of a binary astrophysical source. From now on, we shall use Geometric units $c=G=1$; in such units $1\msun=4.927\times10^{-6}$s and $1$pc$=9.7\times10^7$s. For sources at cosmological distances (i.e., with non negligible redshift $z$), all the following equations are valid if all the masses are replaced by their redshifted counterparts (e.g., $M\rightarrow M_z=M(1+z)$), the distance $D$ is taken to be the luminosity distance (i.e., $D\rightarrow D_L=D(1+z)$) and $f$ is kept to be the {\it observed} GW frequency (the frequency in the source emission restframe is $f(1+z)$). Keeping this in mind, consider a binary system of masses $M_2<M_1$, mass ratio $q=M_2/M_1$ and total mass $M=M_1+M_2$, in circular orbit at a Keplerian frequency $f_K$ at a distance $D$ to the observer. A detailed calculation (in the quadrupole approximation) shows that the system emits a monochromatic wave at a frequency $f=2f_K$, with inclination--polarization averaged GW strain given by \cite{thorne87}:
\begin{equation}
h(f)=\sqrt{\frac{32}{5}}\frac{{\cal M}^{5/3}(\pi f)^{2/3}}{D},
\label{haverage}
\end{equation}
where we introduced the chirp mass ${\cal M}=(M_1M_2)^{3/5}/(M_1+M_2)^{1/5}$. For a pair of Schwarzschild BHs, the maximum frequency of the wave is emitted at the innermost stable circular orbit (ISCO) and can be written as:
\begin{equation}
f_{\rm ISCO}=(\pi6^{3/2}M)^{-1},
\label{fisco}
\end{equation}
where $M_6=M/10^6\msun$. GWs carry away energy from the system, with total luminosity given by: 
\begin{equation}
L_{\rm gw}=\frac{dE_{\rm gw}}{dt}=\frac{32}{5}(\pi{\cal M}f)^{10/3}.
\label{lum}
\end{equation}
Equating the energy loss to the shrinking of the binary semimajor axis $a$,
\begin{equation}
\frac{1}{E}\frac{dE_{\rm gw}}{dt}=-\frac{1}{a}\frac{da}{dt},
\label{dedt}
\end{equation}
and converting $a$ into frequency using Kepler's law yields
\begin{equation}
\frac{df}{dt}=-\frac{64}{5}\pi^{8/3}{\cal M}^{5/3}f^{11/3}.
\label{dfdt}
\end{equation}
The integral of equation (\ref{dfdt}) from $f_0$ to $f_{\rm ISCO}$ defines the remaining lifetime of a binary emitting at a frequency $f_0$ before its final coalescence.
 
Putative eccentricity plays an important role in the evolution of the binary and in the emitted GWs. In this case the luminosity, at a fixed MBHB semimajor axis, is boosted to $L_{\rm gw}=L_{\rm gw,c}F(e)$, where $L_{\rm gw,c}$ is given by equation (\ref{lum}) and 
\begin{equation}
F(e)= (1-e^{2})^{-7/2}\left(1+\frac{73}{24}e^{2}+\frac{37}{96}e^{4}\right).
\label{Fe}
\end{equation}
Accordingly, the evolution of the binary orbit is a factor $F(e)$ faster than in the circular case. The energy is radiated in form of a rather complicated GW spectrum, covering the spectral range $nf_k$, where $n$ is an integer index (see Section \ref{sec4.2} for more details). In particular, the emission is stronger close to the binary periastron (there, the acceleration is larger, and so is the derivative of the quadrupole moment of the source), which leads to efficient circularization according to 
\begin{equation}
\frac{de}{dt} =  -\frac{304}{15}\frac{M_1 M_2 M}{a^4 (1-e^2)^{5/2}}e\left( 1 + \frac{121}{304}e^2 \right).
\end{equation}
Examples of GW driven circularization are illustrated in the MBHB evolutionary paths shown in panels 'a3' and 'b3' of figure \ref{fig2}.

Normalizing equation (\ref{haverage}) to typical astrophysical MBHB values gives a strain of 
\begin{equation}
h\approx 2\times10^{-18}D_9^{-1}{\cal M}_6^{5/3}f_{-4}^{2/3},
\label{strainbin}
\end{equation}
where $D_9=D/10^9$pc and $f_{-4}=f/10^{-4}$Hz. If we require (somewhat arbitrarily) a coalescence timescale $<T_9=T/10^{9}$yr, the integral of equation (\ref{dfdt}), together with equation (\ref{fisco}) implies a relevant frequency range
\begin{equation}
f_{\rm min}=3.54\times10^{-8} T_9^{-3/8}M_6^{-5/8}\,{\rm Hz}<f<f_{\rm ISCO}=4.4\times10^{-3}M_6^{-1}\,{\rm Hz}.
\label{frange}
\end{equation}
If we now estimate the frequency change in an observation time $T$ as $\Delta{f}\approx\dot{f}T$, we find
\begin{equation}
\Delta{f}= 5\times10^{-4} {\cal M}_6^{5/3}f_{-4}^{11/3}T_0 \,{\rm Hz},
\label{fdotrange}
\end{equation}
where now $T_0=T/10^{0}$yr. Equations (\ref{strainbin}), (\ref{frange}) and (\ref{fdotrange}) define the properties of typical MBHB signals. A light $10^5\msun$ binary spans a frequency range of $10^{-7}-10^{-2}$Hz, in a Gyr before the coalescence. For a source at 1Gpc, the strain is $h\approx2\times10^{-19}$ at $10^{-3}$Hz, and $\Delta{f}\gg f$, implying a chirping signal rapidly sweeping through the frequency band during a putative observation. On the other hand, a massive $10^9\msun$ binary covers a frequency range of $10^{-10}-10^{-6}$Hz, in a Gyr before the coalescence. For a source at 1Gpc, the strain is $h\approx5\times 10^{-16}$ at $10^{-8}$Hz, and $\Delta{f}\approx10^{-13}$ Hz, implying a non evolving, monochromatic source. It is likely that many such sources accumulates at these low frequency, resulting in an incoherent superposition of monochromatic waves. MBHBs are therefore loud primary GW targets in the nano-Hz--milli-Hz regime, and they can manifest themselves both as rapidly chirping signals (at milli-Hz frequencies) as well as incoherent superposition of monochromatic waves (at nano-Hz frequencies).

\section{Massive black hole binaries}
\label{sec3}
The mechanism responsible for the formation of the first seed BHs is not well understood. These primitive objects started to form at the onset of the cosmic dawn, around $z\sim 20$, according to current cosmological models \cite{tegmark97}. At an epoch of $z\sim 30-20$, the earliest stars formed in small, metal-poor protogalactic halos may have had masses exceeding 100$\msun$ \cite{bromm99}, ending their lives as comparable stellar mass BHs, providing the seeds that would later grow into MBHs \cite{mr01}. However, as larger, more massive and metal enriched galactic discs progressively formed, other paths for BH seed formation became viable (see \cite{volo10} for a review). Global gravitational instabilities in gaseous discs may have led to the formation of quasi-stars of $10^3-10^6\msun$ that later collapsed into seed BHs \cite{begelman06}. Alternative scenarios are the collapse of massive stars formed in run-away stellar collisions in young, dense star clusters \cite{devecchi09} or the collapse of unstable self-gravitating gas clouds in the nuclei of gas-rich galaxy mergers at later epochs \cite{mayer10}. Thus, the initial mass of the seeds remains one of the largest uncertainties in the present theory of MBH formation.
\begin{figure}
\centering
\includegraphics[width=4.0in]{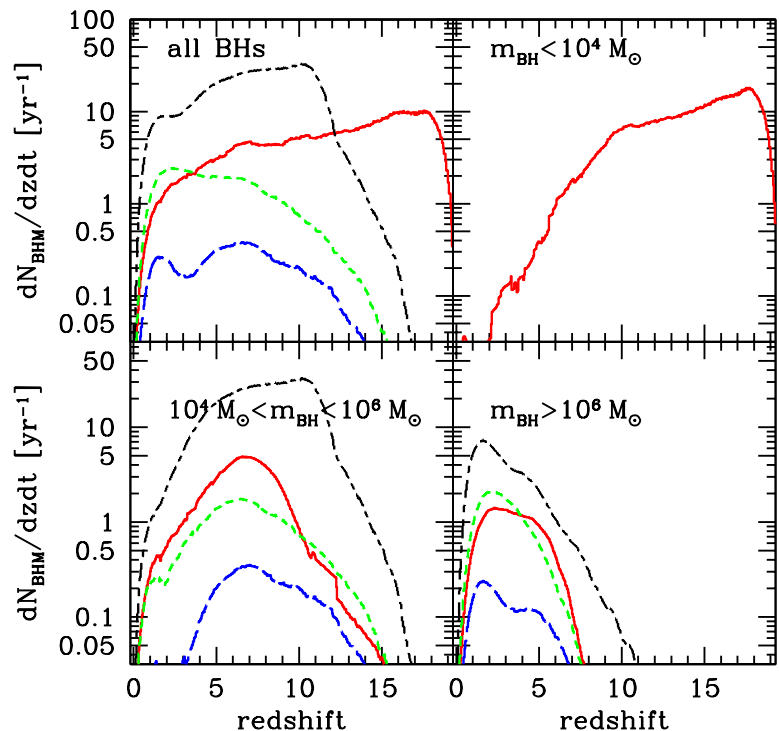}
\caption{Differential MBHB merger rate as a function of redshift for different seed formation scenarios. Adapted from \protect\cite{sesana07}.} 
\label{fig1}
\end{figure}
However, once formed, these seed BHs inevitably took part in the hierarchical structure formation process, growing along the cosmic history through a sequence of mergers and accretion episodes \cite{kh00,vhm03}. Figure \ref{fig1} shows examples of expected MBHB merger rates as a function of redshift for a sample of selected seed BH models. The uncertainty is large, with numbers ranging from ten to several hundreds events per year. Multiplying by the Hubble time and dividing by the number of galaxies within our Hubble horizon ($\approx10^{11}$), figures imply that each galaxy experienced few to few hundred mergers in its past life, placing mergers among the crucial mechanisms in galaxy evolution.

\subsection{Mass and spin evolution}
\label{sec3.1}
Astrophysical BHs are extremely simple objects, described by two quantities only, namely their mass, $M_{\rm BH}$, and angular momentum, ${\bf S}${\footnote{We use boldface to express vectors and standard font to express their magnitude.}}. The magnitude of the latter can be expressed by the dimensionless parameter $a=S/S_{\rm max}=cS/GM_{\rm BH}^2${\footnote{Not to be confused with the binary semimajor axis, it will be clear case by case when we refer to one or the other.}. By definition $0\leq a\leq 1$. Along the cosmic history, MBH mass and spin inevitably evolve according to three principal evolution mechanisms: (i) merger with other MBHs, (ii) episodic accretion of compact objects, disrupted stars, or gas clouds, and (iii) prolonged accretion of large supplies of gas via accretion disks. As shown by \cite{blandford03}, coalescences of MBHs with random spin directions result in a broad remnant spin distribution; in particular highly spinning MBHs tend to spin-down. Despite the important of MBH-MBH mergers,
The dominant role in the mass and spin evolution of MBHs can be attribute to accretion. Continuous Eddington limited accretion implies an exponential mass growth $M_{\rm BH}(t)=M_{\rm BH}(0){\rm exp}\left(\frac{1-\epsilon}{\epsilon}\frac{t}{t_{\rm Edd}}\right)$, where $t_{\rm Edd}=0.45$Gyr and $\epsilon$ is the mass-radiation conversion efficiency ($0.06<\epsilon<0.4$ for $0<a<1$). If this happens in a coherent fashion through, e.g., a thin disk \cite{ss73} , an initially Schwarzschild BH becomes maximally spinning after accreting an amount of mass of the order of $\sqrt{6}M_{\rm BH}(0)$ \cite{thorne74}. However high spins imply $\epsilon>0.3$, considerably slowing down the mass growth, making it impossible to produce a MBH of $>10^9\msun$ at $z=7$ (i.e. in $<10^9$yrs). The problem is avoided if mass is accreted in a series of small incoherent packets (chaotic accretion \cite{king05}). In this case, depending on the angular momentum of the accreted material, the MBH is spun up or down, performing a random walk in spin magnitude that keep it close to zero. However this is true only if the angular momentum direction of the packets is nearly isotropically distributed on the sphere. Real galaxies usually show large coherent gas structures, and a significant amount of rotation (see, e.g., \cite{kassin12,fabricius12}). If the spin vectors of the accreting packets have, on average, a preferential direction (i.e., they angular momenta do not sum up to zero), then the spin evolution is more complicated, and high spin values might still be preferred \cite{pallini13}, as shown in the left panel of figure \ref{fig2}. Rapid mass growth is difficult to reconcile with measurements of high spins (although the latter involve galaxies in the local Universe \cite{reynolds13}), and the requirement of high spins to power energetic relativistic jets in many theoretical models (e.g., \cite{bz77}), and a complete joint understanding of the MBH mass and spin evolution is still missing.
 
\subsection{Massive black hole binary dynamics}
\label{sec3.2}
MBHs become loud sources of GWs when they are in bound, sub--pc binaries. Forming after galaxy mergers, those binaries sit at the center of the stellar bulge of the remnant, possibly surrounded by massive gas inflows triggered by dynamical instabilities related to the strong variations of the gravitational potential during the merger episode \cite{mihos96}. The interaction with the environment imprints distinctive signatures in the binary orbital elements and in the individual spins of the holes. We will see later how this information can be recovered by GW observation, which will therefore allow to directly probe the complex physics underlying the evolution of these spectacular objects.

\begin{figure*}
\centering
\vspace*{-0.0cm}
\begin{tabular}{cc}
\includegraphics[width=2.8in,clip=true]{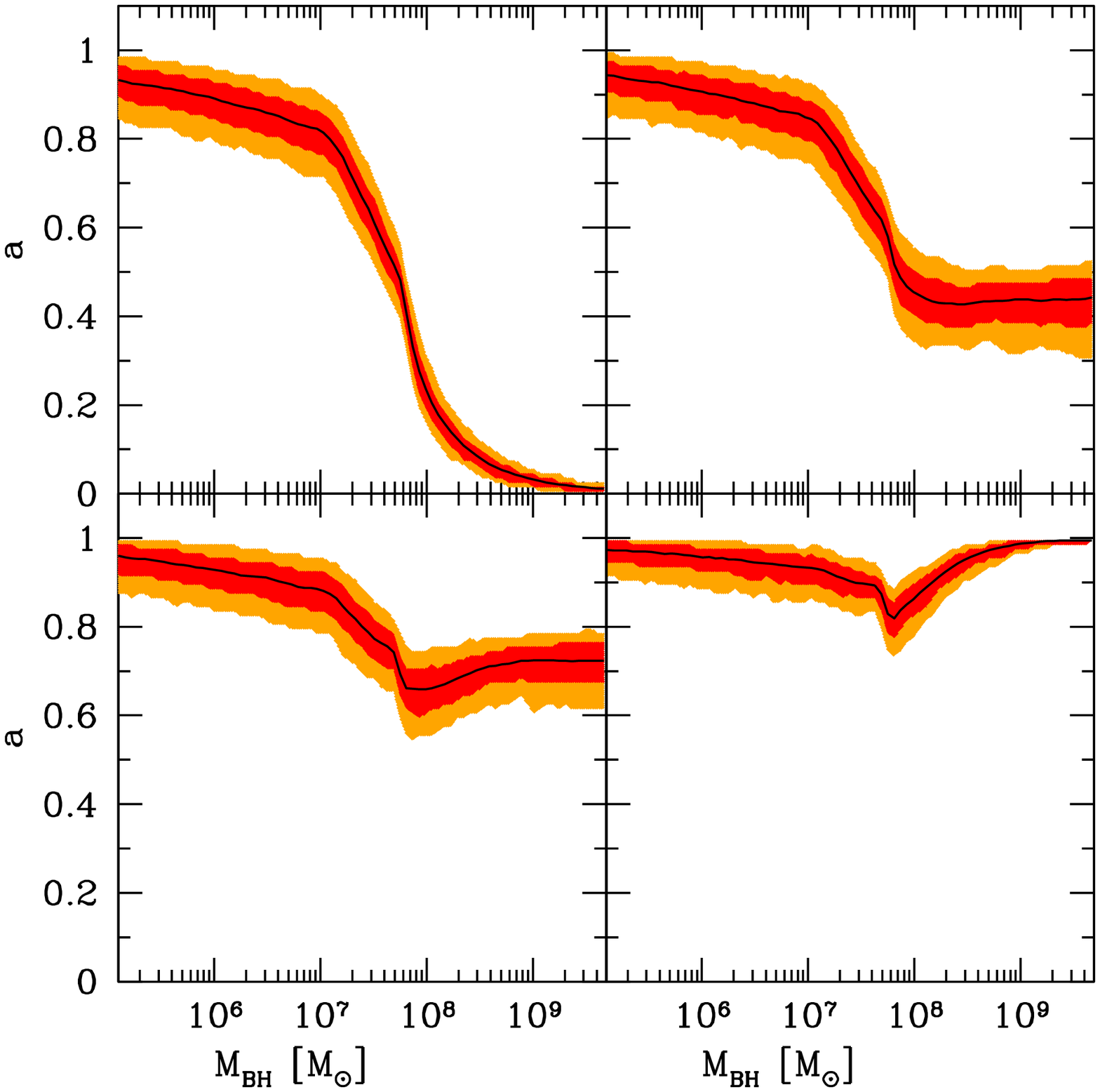}&
\includegraphics[width=3.0in,clip=true]{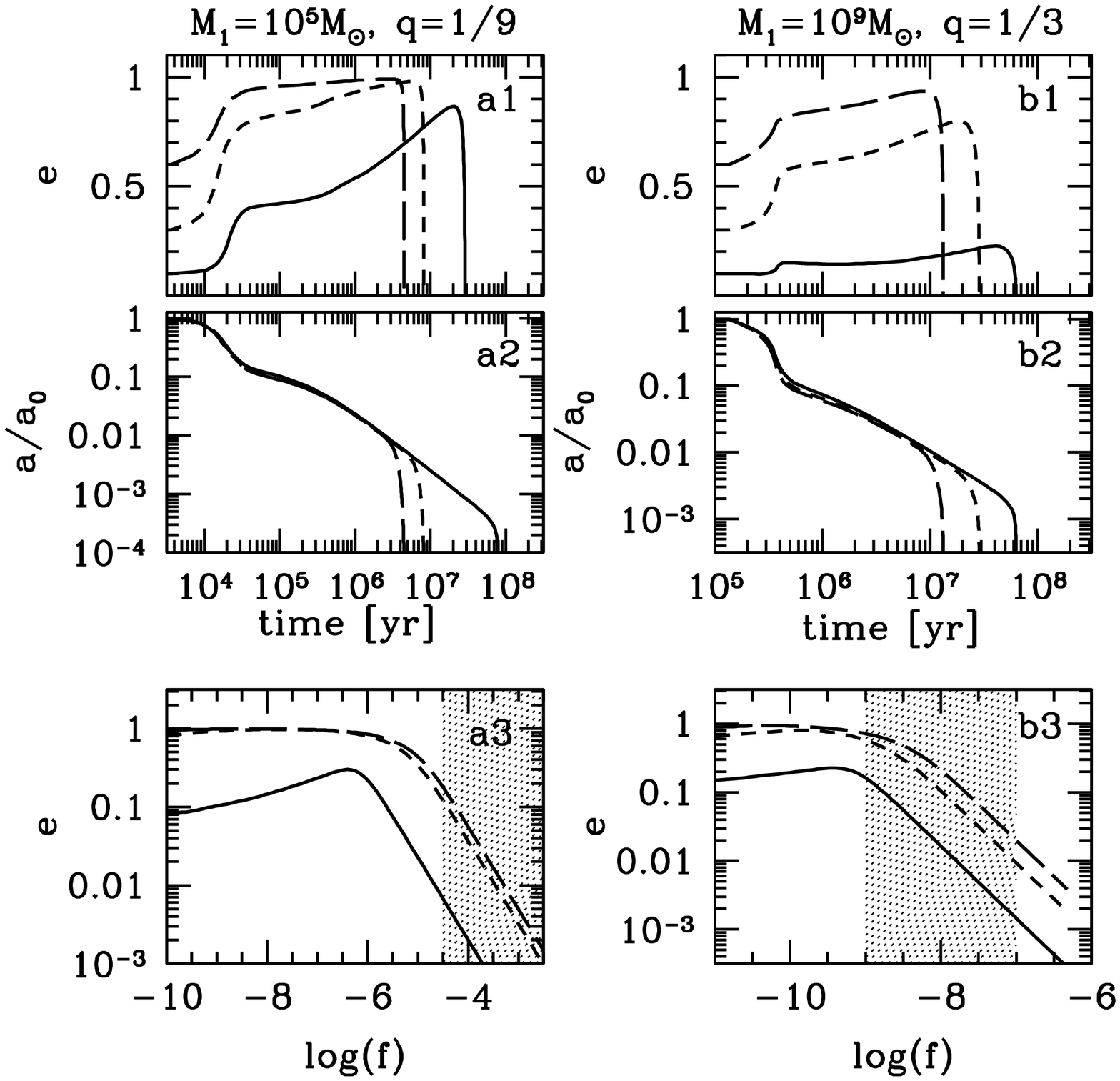}
\end{tabular}
\caption{{\it Left panel}: spin evolution of a MBH accreting incoherent packets of gas of mass $10^{5}\msun$. In this experiment, a parameter $F$ defines the fraction of events in the southern hemisphere (defined with respect to the orientation of the MBH spin). $N\times(1-F)$ accretion events are then isotropically distributed in the northern hemisphere, and $N\times F$ in the southern hemisphere. The upper left, upper right, lower left, and lower right panels refer to $F = 0.5$ (isotropic distribution on the sphere)$, 0.25, 0.125$, and $0$, respectively. The black line refers to the mean over 500 realizations. Red and orange shaded areas enclose intervals at 1-$\sigma$ and 2-$\sigma$ deviations, respectively (from \protect\cite{pallini13}). {\it Right  panel}: examples of MBHB evolutionary tracks in stellar environments \protect\cite{sesana10}. Shown are the evolution of the eccentricity and binary semimajor axis in time ('1' and '2' panels) and the evolution of the eccentricity versus the orbital frequency ('3' panels). The very high eccentricities achieved in the stellar driven phase imply non negligible eccentricities in the milli-Hz regime (probed by future space based interferometers like eLISA, shaded area in panel 'a3'), and potentially extremely high eccentricities in the nano-Hz regime (targeted by PTAs, shaded area in panel 'b3').}  
\label{fig2}
\end{figure*}

\subsubsection{Stellar driven binaries.}
Ignoring technical details related to the 'loss cone evolution' (see D. Merritt contribution to this issue), a background of stars scattering off the binary drives its semimajor axis evolution according to the equation \cite{quinlan96}
\begin{equation}
\frac{da}{dt} = \frac{a^2G\rho}{\sigma}H,
\label{adotstar}
\end{equation}
where $\rho$ is the density of the background stars, $\sigma$ is the stellar velocity dispersion and $H$ is a numerical coefficient of order 15. 
The eccentricity evolution in stellar environments has been tackled by several authors by means of full N-body simulations. In general, equal mass, circular binaries tend to stay circular or experience a mild eccentricity increase \cite{mms07}, while binaries that form already eccentric, or with $q\ll 1$ (regardless of their initial eccentricity) tend to grow more eccentric \cite{mat07,preto11}, in reasonable agreement with the prediction of scattering experiments \cite{quinlan96,sesana06}. The interaction with stars is unlikely to significantly affect the individual spins of the holes. Therefore, star driven binaries are expected to grow to quite high eccentricities, while the spins of the individual holes are likely randomly oriented. 

\subsubsection{Gas driven binaries.}
In the case of circumbinary disks, the detailed evolution of the system depends on the complicated and uncertain dissipative physics of the disk itself. The simple case of a coplanar prograde circumbinary disk admits a selfconsistent, non stationary solution that was derived by \cite{ipp99}. In this case, the binary semimajor axis evolution can be approximated as \cite{ipp99,haiman09}
\begin{equation}
\frac{da}{dt} = \frac{2\dot{M}}{\mu}(aa_0)^{1/2}.
\label{adotgas}
\end{equation}
Here, $\dot{M}$ is the mass accretion rate at the outer edge of the disk, $a_0$ is the semimajor axis at which the mass of the unperturbed disk equals the mass of the secondary MBH, and $\mu$ is the reduced mass of the binary. 
In the circumbinary disk scenario, eccentricity excitation has been seen in several simulations \cite{armitage05,cuadra09}. In particular, the existence of a limiting eccentricity  $e_{\rm crit} \approx 0.6 - 0.8$ has been found in \cite{roedig11}, 
in the case of massive selfgravitating disks. If the accretion flow is coherent, and ${\bf L}_{\rm disk}$ and the spins ${\bf S}_i$ of the two MBHs are misaligned, the Bardeen-Petterson effect \cite{bp75} will act to align ${\bf S}_i$ to ${\bf L}_{\rm disk}$ in a very short timescale ($t_{\rm align}\ll t_{\rm acc}\sim10^8$yr \cite{scheuer96,perego09}). Therefore, in gaseous rich environments, mildly eccentric binaries might be the norm, and the MBH individual spins tend to align with the orbital angular momentum.

Compared to the GW driven case, $(da/dt)_{\rm gw}\propto a^{-3}$, equations (\ref{adotstar}) and (\ref{adotgas}) have a very different (milder and positive) $a$ dependence. Therefore, equating  equations (\ref{adotstar}) and (\ref{adotgas}) to $(da/dt)_{\rm gw}$ gives the transition frequency between the external environment driven and the GW driven regimes. For typical astrophysical systems one gets: 
\begin{eqnarray}
f_{{\rm star/GW}}\approx 1.2\times10^{-7}M_6^{-7/10}q^{-3/10} {\rm Hz}\nonumber\\
f_{{\rm gas/GW}}\approx 1.6\times10^{-7}M_6^{-37/49}q^{-69/98} {\rm Hz}.
\label{decoup}
\end{eqnarray}
We therefore see that very massive nano-Hz MBHBs might still be influenced by their environment, and therefore have high eccentricities ($e>0.5$, see panel 'b3' in figure \ref{fig2}). Even though GW emission efficiently circularizes binaries (see Section \ref{sec2}), systems in the milli-Hz range can still retain substantial residual eccentricities ($e>0.01$, see panel 'a3' in figure \ref{fig2}).

\section{MBHB waveforms}
\label{sec4}
Having introduced the basics of GW emission from a binary system in Section 2, we turn now in some more detail to the gravitational waveform modeling. In particular we show how eccentricity and spins affect the detectable GW signal and we describe the basic theory of information recovery, that enables us to dig out the parameters of the source from the detected waveform. 

\subsection{The stages of the binary coalescence}
\label{sec4.1}
\begin{figure}
\centering
\includegraphics[width=6.0in]{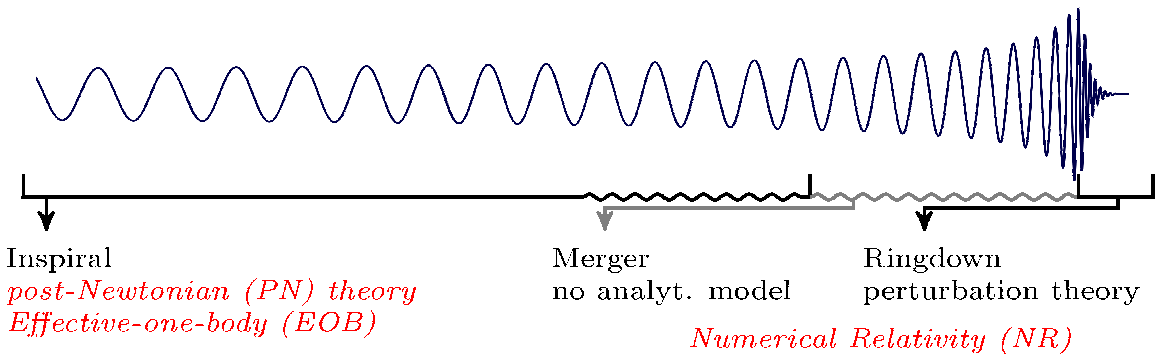}
\caption{Example of GW signal from two coalescing (circular, non spinning) BHs as a function of time. The different approximation techniques and their range of validity are indicated. Wavy lines illustrate the regime close to merger where analytical methods have to be bridged by NR (courtesy of F. Ohme \protect\cite{ohme12}).} 
\label{fig3}
\end{figure}
The evolution of MBHBs is customarily divided into three phases: inspiral, merger, and ring-down \cite{flanagan98}. The inspiral is a relatively slow, adiabatic process. Different techniques have been employed to describe this stage, ranging from classic Post Newtonian (PN) expansions of the energy--balance equation \cite{blanchet06}, to non--adiabatic resummed methods in which the equations of motion are derived from an effective one body (EOB) relativistic Hamiltonian \cite{buonanno99}. A detailed description of such methods is beyond the scope of this article, and an excellent overview can be found in \cite{buonanno03}. The inspiral is followed by the dynamical coalescence, in which the MBHs plunge and merge together, forming a highly distorted, perturbed remnant. 
At this stage, all analytical approximations break down, and the system can only be described solving by directly the Einstein equation using numerical simulations \cite{pretorius05,campanelli06,baker06}. The distorted remnant settles into a stationary Kerr BH as it rings down, by emitting gravitational radiation. This latter stage can be, again, modeled analytically using BH perturbation theory \cite{berti06}. 
An example of the full waveform with the identification of the various stages is given in figure \ref{fig3}. 

In recent years there has been a major effort in constructing accurate waveforms inclusive of all three phases. "Complete" waveforms can be designed by stitching together analytical PN waveforms for the early inspiral with a (semi)phenomenologically described merger and ring-down phase  calibrated against available numerical data (known as PhenomB-PhenomC waveforms, \cite{santamaria10}). Alternatively, complete waveforms can be constructed within the EOB formalism by adding free parameters to be calibrated against NR simulations and by attaching a series of damped sinusoidals describing the ringdown (known as EOBNR waveforms \cite{damour08,buonanno09}). A detailed overview is given in \cite{ohme12}. What is relevant to our discussion is that the full evolution of MBHBs can be tackled with a combination of analytical and numerical methods, and accurate waveforms encoding all the parameters of the system can be computed. In the following we concentrate on the inspiral signal only, which is the richest in terms of encoded information.
 
\subsection{The adiabatic inspiral: impact of eccentricity and spin}
\label{sec4.2}
For circular binaries, the evolution of the adiabatic inspiral is completely determined by the energy-balance equation that relates the derivative of the energy function ${\cal E}$ to the gravitational flux ${\cal F}$ radiated away {\footnote{For eccentric binaries, an angular momentum balance equation must also be imposed to compute the evolution in eccentricity.}} 
\be
\frac{d{\cal E}(v)}{dt}=-{\cal F}(v)
\label{flux}
\ee
from which we may derive the binary acceleration and phase evolution as:
\be
\frac{d\Phi}{dt}=\frac{2v^3}{M},\,\,\,\,\,\,\,\,\,\,\frac{dv}{dt}=-\frac{{\cal F}(v)}{Md{\cal E}(v)/dv}.
\label{flux}
\ee
For orbital velocities $v\ll c$, ${\cal E}$ and ${\cal F}$ can be expanded in powers of $v^{2n}$ to a given order in $n$. This results in a corresponding expansion for the binary acceleration of the form \cite{kidder95}:
\be
{\bf a} = {\bf a}_{N} + {\bf a}_{PN} + {\bf a}_{SO} + {\bf a}_{2PN} + {\bf a}_{SS} + {\bf a}_{RR},
\label{agen}
\ee
where ${\bf a}_N$, ${\bf a}_{PN}$, and ${\bf a}_{2PN}$ are the Newtonian, (post)1-Newtonian, and (post)2-Newtonian contributions to the equations of motion, ${\bf a}_{RR}$ is the contribution due to the radiation-reaction force, and ${\bf a}_{SO}$ and ${\bf a}_{SS}$ are the spin-orbit and spin-spin coupling contributions. This is reflected in the radiated wave, that can be similarly expanded in the form:
\be
h^{ij}=\frac{2\mu}{D}\left[Q^{ij} + P^{0.5}Q^{ij} + PQ^{ij} + PQ^{ij}_{SO} + P^{1.5}Q^{ij} + P^{1.5}Q^{ij}_{SO}+P^{2}Q^{ij}_{SS}\right]
\label{hgen}
\ee
where $Q^{ij}$ is the standard quadrupole moment of the source, $i,j=1,2,3$ define the spatial components of the perturbation tensor, and the subscripts have the same meaning as in equation (\ref{agen}). Choosing the appropriate orthonormal radiation frame, $h^{ij}$ can be written as two independent polarizations only; $h_+$ and $h_\times$. To the leading quadrupole order, in the circular case one obtains the familiar form
\ba
h_+(t) = 2 \frac{{\cal M}^{5/3}}{D}\,\left[\pi f(t)\right]^{2/3}(1 + \cos^2 \iota)\cos\Phi(t)\nonumber\\
h_\times(t) = 2 \frac{{\cal M}^{5/3}}{D}\,\left[\pi f(t)\right]^{2/3}2\sin^2\iota\cos\Phi(t)
\label{hquadcirc}
\ea
where $\iota$ (usually referred as inclination) is the angle defined by the line of sight with respect to the orbital angular momentum vector, $\Phi(t) = 2\pi\int^t f(t') dt'$, and $f=2f_K$ as defined in Section \ref{sec2}.

The eccentricity $e$ enters directly in the computation of $Q^{ij}$ since it affects the velocity ${\bf v}$ of the MBHs along the orbit. In fact $e$ affects the computation of $h^{ij}$ at all orders, starting from the simple quadrupole term, by "splitting" each polarization amplitude $h_+(t)$ and $h_\times(t)$ into harmonics according to (see, e.g., equations (5-6) in \cite{wvk08} and references therein):
\ba
h^{+}_n(t) = A \Bigl\{-(1 + \cos^2\iota)u_n(e) \cos\left[\frac{n}{2}\,\Phi(t) + 2 \gamma(t)\right]\nonumber \\
\,\,\,\,\,\,\,\,-(1 + \cos^2\iota) v_n(e) \cos\left[\frac{n}{2}\,\Phi(t) - 2 \gamma(t)\right]
+ \sin^2\iota\, w_n(e) \cos\left[\frac{n}{2}\,\Phi(t)\right] \Bigr\},
\label{e:h+}\nonumber\\
h^{\times}_{n}(t) = 2 A \cos\iota \Bigl\{u_n(e) \sin\left[\frac{n}{2}\,\Phi(t) + 2 \gamma(t)\right] 
+ v_n(e) \sin\left[\frac{n}{2}\,\Phi(t) - 2 \gamma(t)\right] \Bigr\}\,.
\label{e:hx}
\ea
The amplitude coefficients $u_n(e)$,  $v_n(e)$, and $w_n(e)$ are linear combinations of the Bessel functions of the first kind $J_{n}(ne)$, $J_{n\pm 1}(ne)$ and $J_{n\pm 2}(ne)$, and $\gamma(t)$ is an additional precession term to the phase given by $e$. For $e\ll 1$, $|u_n(e)| \gg |v_n(e)|\,,|w_n(e)|$ and we recover the circular limit given by equation (\ref{hquadcirc}). As GW emission tends to decrease eccentricity, this is likely to be mostly important at large separations (i.e., $f_K\ll f_{\rm ISCO}$, see panels 'a3' and 'b3' in figure \ref{fig2}). An example of an eccentric waveform is shown in the upper panel of figure \ref{fig4}.
\begin{figure}
\centering
\includegraphics[width=4.0in]{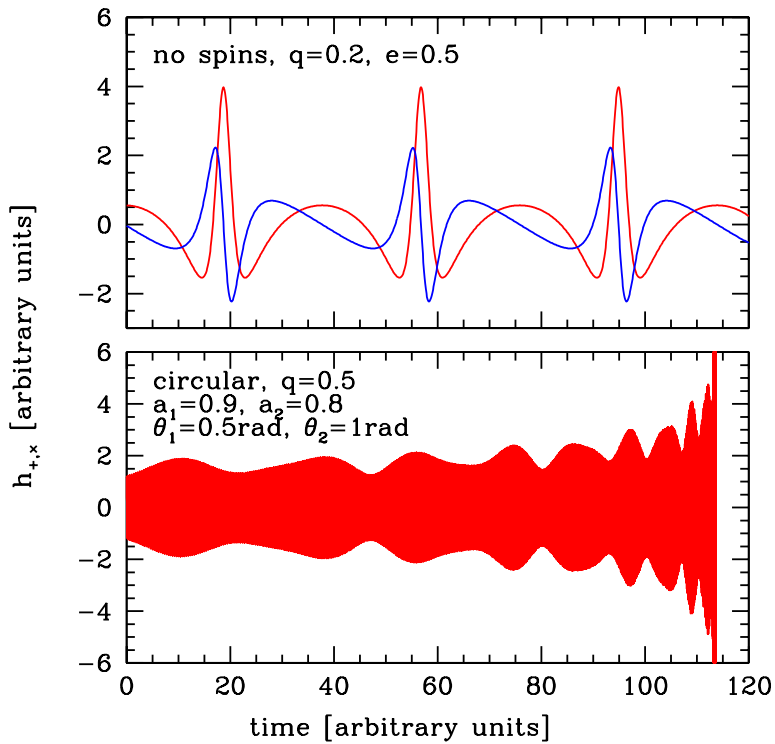}
\caption{Examples of waveforms from eccentric and spinning binaries. In the top panel we show a zoom-in of three wave cycles highlighting the peculiar amplitude-phase relation of a mildly eccentric binary ($e=0.5$); $h_+$ is in red and $h_\times$ is in blue. In the bottom panel we show the last several thousand cycles of a spinning precessing system, the amplitude modulation given by the orbital plane precession is evident (courtesy of A. Petiteau).} 
\label{fig4}
\end{figure}

Turning now to spins, equation (\ref{agen}) shows that spins enter as higher order corrections in the computation of the acceleration and, consequently, of the waveform. Therefore they become important only when $v/c$ is substantial, significantly affecting the signal only for $f>0.01 f_{\rm ISCO}$. Such corrections generate additional small terms in the phase evolution, but most importantly, cause the orbital angular momentum  ${\bf L}$ to precess around the total conserved angular momentum ${\bf J}={\bf L}+{\bf S}_1+{\bf S}_2$ (in which ${\bf S}_1$ and ${\bf S}_2$ the spins of the two MBHs) according to a precession equation, that to the leading order reads \cite{vecchio04}
\be
\dot{{\bf L}} = \left[\frac{1}{a^3}\left(2 + \frac{3}{2}q\right){\bf J}\right]\times {\bf L}
\label{precession}
\ee
(in this case $a$ is the binary separation). The typical precession timescale is given by 
\be
T_p=\frac{L}{\dot{L}}\approx M^{1/3}\mu^{-1}\left[2+\frac{3}{2}q\right]^{-1}(\pi f)^{5/3}\approx 2.5\times10^{3}\pi^{5/3}M_6^{1/3}\left(\frac{f}{f_{\rm ISCO}}\right)^{-5/3}
\label{prectime}
\ee
where we used ${\bf J} \approx {\bf L}$, and we assumed equal mass binaries in the last equality. It is clear that at, say, $f=10^{-3}f_{\rm ISCO}$, $T_p$ is of the order of several tens of years, making precession effects negligible. An example of waveform modulated by plane precession in the late inspiral is  shown in the lower panel of figure \ref{fig4}.

The most general detectable signal from a spinning eccentric binary is a function of 17 parameters (some describing the intrinsic properties of the binary and some others related to the relative binary-detector position and orientation): two combination of the redshifted masses, ${\cal M}$ and $\mu${\footnote{As discussed in Section 2, for sources at cosmological distances, the GW depends on the redshifted masses, the intrinsic ones can be extracted by measuring $D_L$ and then measuring $z$ according to a cosmological model, or by obtaining an independent measurement of $z$ through, e.g., the identification of an electromagnetic counterpart to the GW signal.}}; six parameters defining the individual spin vectors, two magnitudes $a_1$ and $a_2$ and four angles; two parameters related to the eccentricity of the orbit, initial eccentricity $e_0$ and an additional angle defining the line of nodes; source inclination with respect to the line of sight, $\iota$; polarization angle, $\psi$; sky location, two angle usually labeled $\theta$ and $\phi${\footnote{Of particular interest is the source sky localization errorbox $\Delta\Omega$, defined, following \cite{cutler98}, in terms of $\theta$ and $\phi$ as: $\Delta \Omega=2\pi\sqrt{({\rm sin}\theta \Delta \theta \Delta \phi)^2-({\rm sin}\theta c^{\theta\phi})^2}$ (according to the notation used in Section \ref{sec4}).}}; luminosity distance $D_L$; initial orbital phase $\Phi_0$; and, depending on the type of signal, initial frequency $f_0$ or time to coalescence $t_c$ (the two are related, in the quadrupole approximation, by equation (\ref{dfdt})). We saw examples of how some of these parameters are imprinted in the waveform, in the following subsection we turn to the problem of how accurately they can be extracted given some signal observation.

\subsection{Parameter extraction}
\label{sec4.3}
We briefly review the basic theory regarding the estimate of the statistical errors that affect the measurements of the source parameters. For a comprehensive discussion of this topic we refer the reader to~\cite{jaynes03}. The data $d$ collected in a detector is given by the superposition of the noise $n$ and a signal $x$ determined by a vector of parameters $\vec{\lambda}$: 
\be
d(t) = n(t) + x(t;\vec{\lambda})\,.
\label{e:da}
\ee
In the following, we make the usual (simplifying) assumption that $n(t)$ is a zero-mean Gaussian and stationary random process characterized by the one-sided power spectral density $S_n(f)$. 
The inference process in which we are interested is how well one can infer the actual value of the unknown parameter vector $\vec\lambda$, based on the data $d$, and any prior information on $\vec\lambda$ available before the experiment. Within the Bayesian framework, one is therefore interested in deriving the posterior probability density function (PDF) $p(\vec\lambda | d)$ of the unknown parameter vector given the data set and the prior information. Bayes' theorem yields
\be
p(\vec\lambda | d) = \frac{p(\vec\lambda)\,p(d|\vec\lambda)}{p(d)}\,,
\label{e:posterior}
\ee
where $p(d|\vec\lambda)$ is the likelihood function, $p(\vec\lambda)$ is the prior probability density of $\vec\lambda$, and $p(d)$ is the marginal likelihood or evidence. Under the assumption of Gaussian noise 
\be
p(d|\vec\lambda) \equiv p[n = d-x(\vec\lambda)] \propto \exp{\left[-\frac{1}{2}(d-x(\vec\lambda)|d-x(\vec\lambda))\right]}
\ee
where the inner product between two functions $(g|h)$ is defined as the integral
\be
(g|h) = 2 \int_{0}^{\infty} \frac{\tilde g^*(f) \tilde h(f) +  \tilde g(f) \tilde h^*(f)}{S_n(f)} df\,,
\label{e:innerxy}
\ee
applied to the Fourier Transform of the functions, e.g.,
\be
\tilde g(f) = \int_{-\infty}^{+\infty} g(t) e^{-2\pi i f t}.
\label{e:tildex}
\ee
In the neighborhood of the maximum-likelihood estimate value $\hat{{\vec \lambda}}$, the likelihood function can be approximated as a multi-variate Gaussian distribution,
\be
p(\vec\lambda | d) \propto p(\vec\lambda)
 \exp{\left[-\frac{1}{2}\Gamma_{ab} \Delta\lambda_a \Delta\lambda_b\right]}\,,
\label{linpdf}
 \ee
where $ \Delta\lambda_a = \hat{\lambda}_a - {\lambda}_a$ and the matrix $\Gamma_{ab}$ is the Fisher information matrix (FIM). Here $a,b = 1,\dots, N$ label the components of $\vec{\lambda}$ (i.e., the parameters defining the shape of the signal), and we have used Einstein's summation convention (and we do not distinguish between covariant and contravariant indices). The FIM is simply related to the derivatives of the GW signal with respect to the unknown parameters integrated over the observation:
\be
\Gamma_{ab} = \left(\frac{\partial x(t; \vec\lambda)}{\partial\lambda_a} \Biggl|\Biggr.\frac{\partial x(t; \vec\lambda)}{\partial\lambda_b}
\right)\,.
\label{e:Gamma_ab_a}
\ee
In terms of the inner product $(.|.)$ the maximal signal-to-noise ratio (SNR) at which a signal can be observed is obtained by matched filtering of the data against a template equal to the waveform signal, $x(\vec\lambda)$. The optimal matched filtering SNR achievable in this way is $(x|x)$. In the limit of large SNR, $\hat{{\vec \lambda}}$ tends to ${{\vec \lambda}}$, and the inverse of the FIM provides a lower limit to the error covariance of unbiased estimators of ${{\vec \lambda}}$, the so-called Cramer-Rao bound. The variance-covariance matrix is simply the inverse of the FIM, and its elements are
\be
\sigma_a^2 = \left(\Gamma^{-1}\right)_{aa},\,\,\,\,\,\,\,\,\,\,\,c_{ab} = \frac{\left(\Gamma^{-1}\right)_{ab}}{\sqrt{\sigma_a^2\sigma_b^2}}\,,
\label{e:cab}
\ee
where $-1\le c_{ab} \le +1$ ($\forall a,b$) are the correlation coefficients. We can therefore interpret $\sigma_a^2$ as a way to quantify the expected uncertainties on the measurements of the source parameters. We refer the reader to~\cite{vallisneri08} and references therein for an in-depth discussion of the interpretation of the inverse of the FIM in the context of assessing the prospect of the estimation of the source parameters for GW observations. 
When combining $\alpha=1,...,N$ different pieces of independent (i.e., having uncorrelated noise) information (for example, by observing several pulsars, or by combining the inspiral and the ringdown portion of a signal), the FIM that characterizes the \emph{joint} observations in equation (\ref{linpdf}) is simply given by the sum of the matrices of all the individual pieces
\be
\Gamma_{ab} = \sum_\alpha \Gamma_{ab}^{(\alpha)}\,,
\ee
and all the theory follows unchanged.

\section{The gravitational wave landscape: observations and scientific payouts}
\label{sec5}
As shown in Section \ref{sec2}, GW emission from MBHBs covers several decades in frequency, ranging from sub-nano-Hz to milli-Hz. As shown in figure \ref{fig5}, this range is (or it will be) covered by multiple probes. The ground-based network of advanced interferometric detectors (three LIGO detectors, VIRGO \cite{aasi13}, and the Kamioka Gravitational wave Detector, KAGRA \cite{somiya12}) and possibly the third-generation Einstein Telescope (ET, \cite{punturo10}) will observe inspiralling binaries up to around few$\times100\msun$. The milli-Hz regime will be the hunting territory of spaced based detectors such as eLISA, whereas PTAs are already probing the nano-Hz portion of the frequency band. In particular, space based interferometers and PTAs will provide a complementary, complete census of the MBHB population throughout the Universe. In this section we focus on the prospects of GW observation in these two bands and on the related scientific payouts.
\begin{figure}
\centering
\includegraphics[width=4.5in]{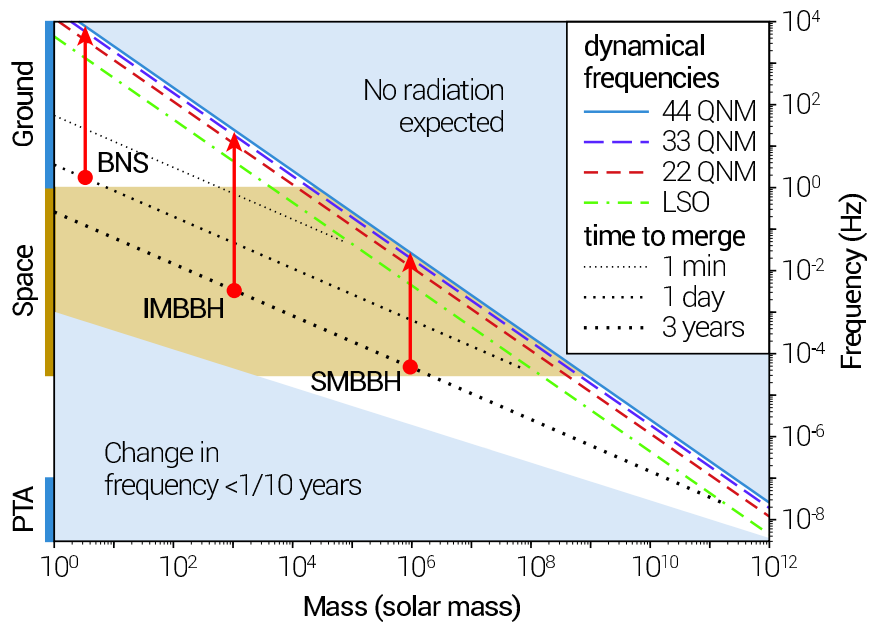}
\caption{The GW landscape; total mass of the binary system versus frequency of the GWs. The left side shows the frequency bands of different probes: ground based detectors, eLISA, and PTAs. The gray shaded region at the top is inaccessible because no system of a given mass can radiate at such high frequencies. The shaded region at the bottom contains detectable sources for which the chirp mass cannot be measured in an observation lasting less than 10 years.
Sloping dotted lines show the three-year, one-day, and one-minute time-to-merger lines. Sloping dashed lines are relevant dynamical frequencies: last stable orbit and the frequencies of ringdown modes of the merged BH. Vertical lines indicate evolutionary tracks of systems of various masses as their orbits shrink and they move to higher frequencies: a binary neutron star (BNS), an intermediate-mass binary BH (IMBBH), and a super massive binary BH (SMBBH) (from \protect\cite{whitepaper13}).} 
\label{fig5}
\end{figure}

\subsection{The milli-Hz regime: science with space based interferometry}
\label{sec5.1}
Space based interferometry will open a revolutionary new window on the Universe. In the following we refer to the eLISA design presented in \cite{whitepaper13} to describe the extraordinary scientific payouts of milli-Hz GW observations. Figure \ref{fig6} highlights the exquisite capabilities of eLISA in covering almost all the mass-redshift parameter space relevant to MBH astrophysics. 
GW observations will catch sources with $M\sim10^4\msun$ at early cosmological times, prior to reionization. A binary with $10^4\lesssim M\lesssim 10^7\msun$ can be detected out to  $z\sim 20$ with a SNR $\ge 10$, making an extensive census of the MBH population in the Universe possible.
\begin{figure}
\centering
\includegraphics[width=4.5in]{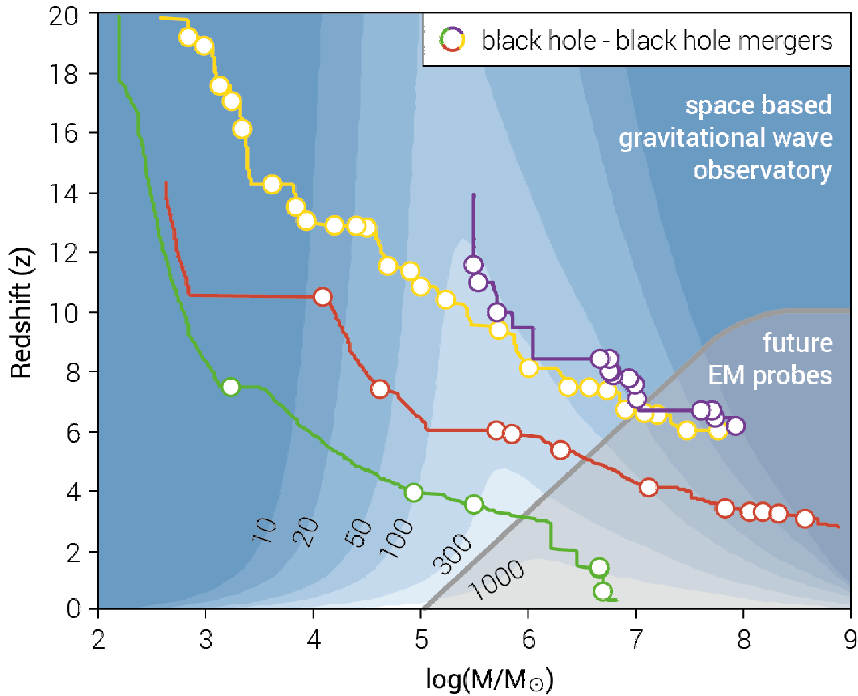}
\caption{Constant-contour levels of the sky and polarization angle averaged SNR for eLISA, for equal mass non-spinning binaries as a function of their total rest frame mass, $M$, and cosmological redshift, $z$. The tracks represent the mass-redshift evolution of selected supermassive BHs: two possible evolutionary paths for a BH powering a $z\approx6$ quasar (starting from a massive seed, blue curve, or from a Pop III seed from a collapsed metal-free star, yellow curve); a typical $10^9\msun$ BH in a giant elliptical galaxy (red curve); and a Milky Way-like BH (green curve). Circles mark BH-BH mergers occurring along the way \protect\cite{volonteri09}. The gray transparent area in the bottom right corner roughly identifies the parameter space for which MBHs might power phenomena that will likely be observable by future electromagnetic probes.} 
\label{fig6}
\end{figure}

As detailed in Section \ref{sec4}, detected waveforms carry information on all the relevant source parameters, including redshifted masses 
and spins of the individual BHs prior to coalescence, the distance to the source and its sky location. 
The left panel of figure \ref{fig7} shows error distributions in the source parameter estimation, for events collected in a meta-catalog of $\sim 1500$ sources, based on state of the art MBH evolution models (see \cite{amaro13a} for details). 
Here circular precessing spinning binary were considered (i.e. waveforms determined by 15 parameters), and "hybrid" waveforms of the PhenomC family were used to evaluate uncertainties based on the FIM approximation, as outlined in the previous section. Individual redshifted masses can be measured with unprecedented precision, i.e. with an error of $0.1\%-1\%$, on both components. The spin of the primary hole can be measured with an exquisite accuracy, to a 0.01-0.1 absolute uncertainty. This precision mirrors the big imprint left by the primary MBH spin in the waveform. The measurement is more problematic for $a_2$ that can be either determined to an accuracy of 0.1, or remain completely undetermined, depending on the source mass ratio and spin amplitude. The source luminosity distance error has a wide spread, usually ranging from being undetermined (but see \cite{sesanalisa9} for possible shortcomings of the FIM approximation in these cases) to a stunning few percent accuracy (note that this is a direct measurement of $D_L$)
GW detectors are full sky monitors, and the localisation of the source in the sky is also encoded in the waveform pattern. Sky location accuracy is typically estimated in the range 10-1000 square degrees{\footnote{These numbers assume a 'single Michelson' (four laser links) configuration for eLISA, the full triangular 'two Michelsons' (six laser links) configuration results in a significant improvement of the estimation of all parameters, in particular luminosity distance and sky location (by 1-2 orders of magnitude).}}.
\begin{figure*}
\centering
\vspace*{-0.0cm}
\begin{tabular}{cc}
\includegraphics[width=2.9in,clip=true]{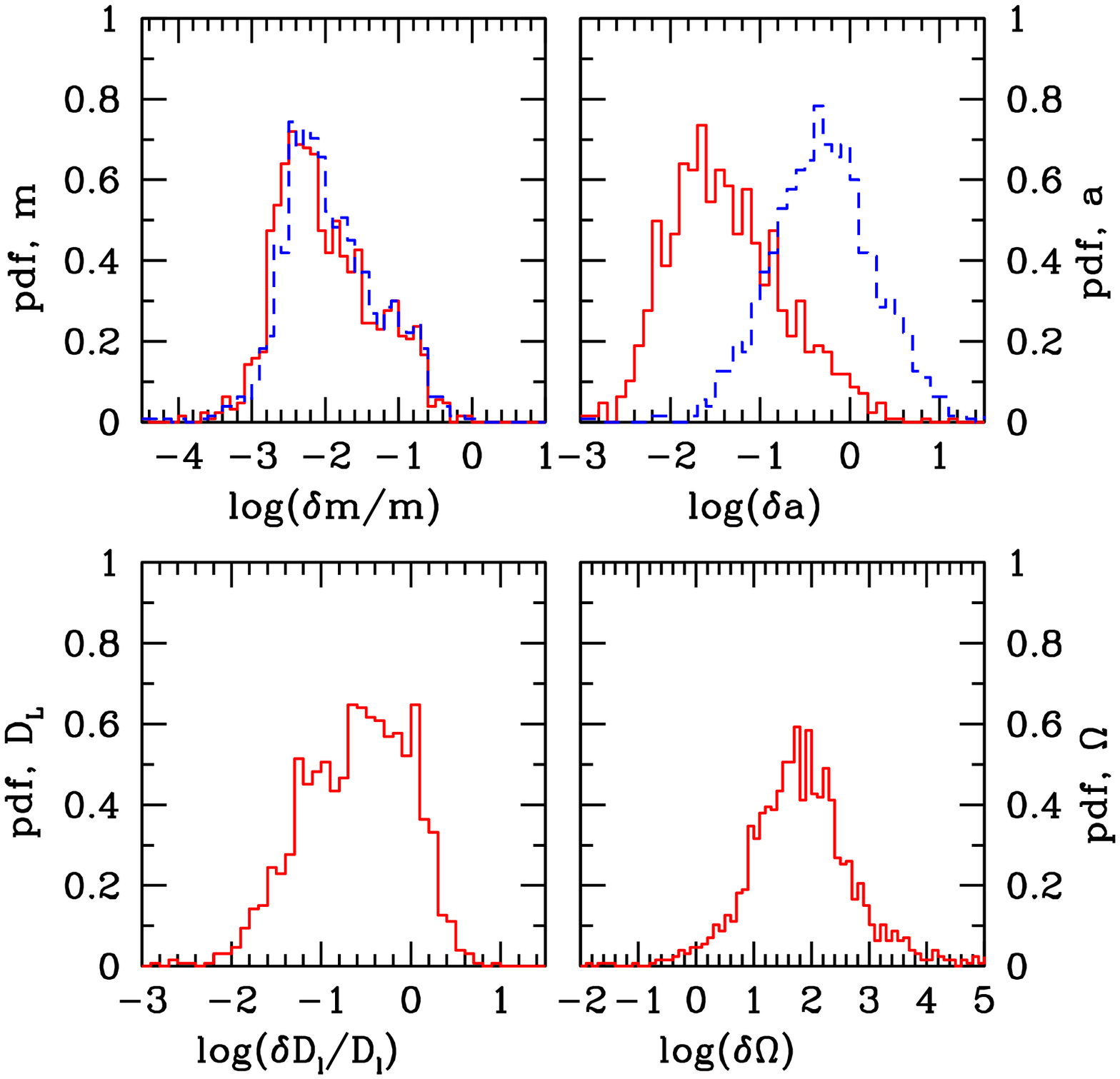}&
\includegraphics[width=2.9in,clip=true]{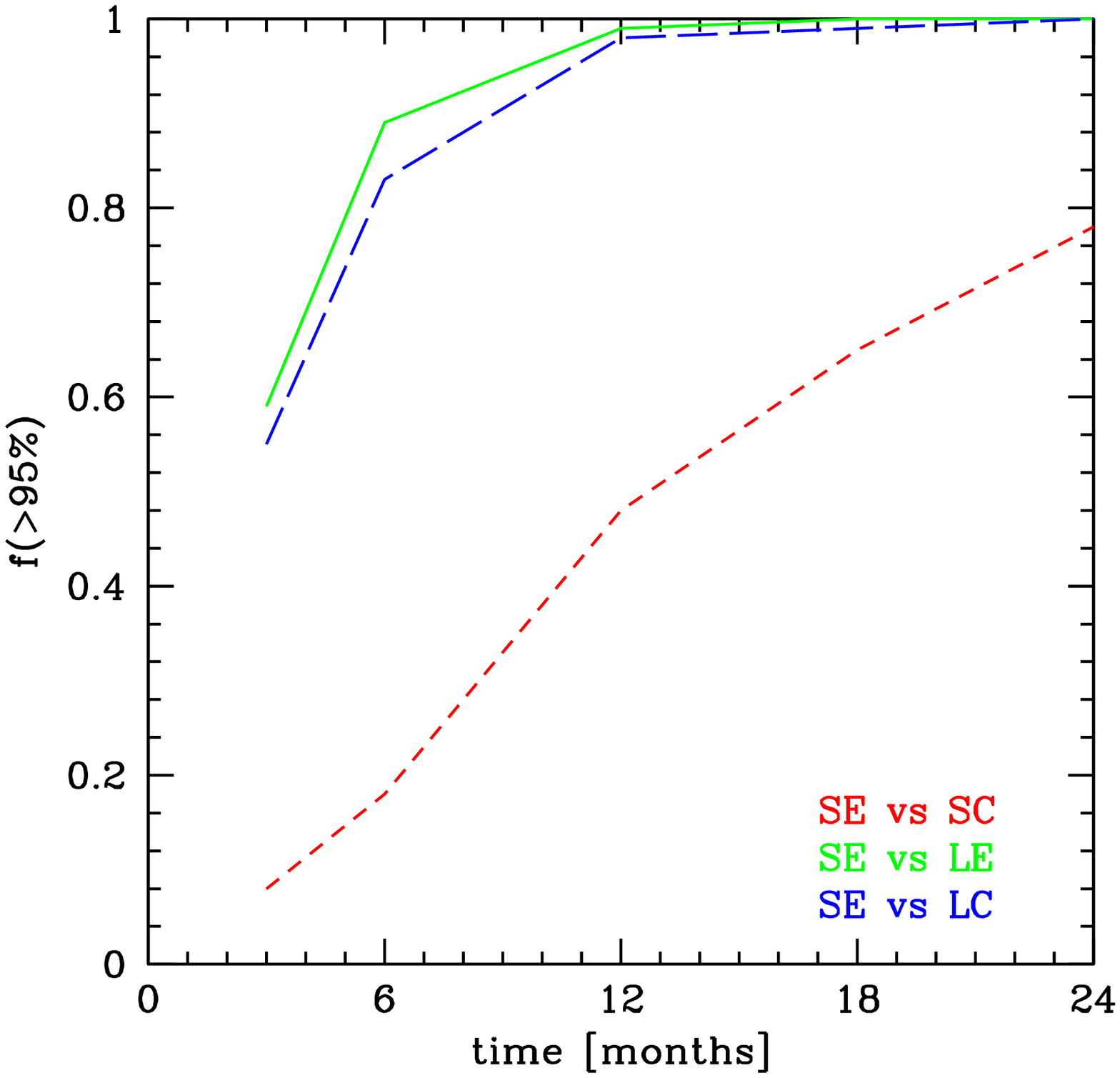}
\end{tabular}
\caption{{\it Left panel}: eLISA parameter estimation accuracy; meta-catalog of MBHBs described in \protect\cite{amaro13a}. Top panels show errors on the redshifted masses (left) and spins (right). Red solid lines are for the primary and blue dashed lines are for the secondary MBH. The bottom panels show the error distribution on the luminosity distance $D_L$ (left), and the sky location accuracy $\Delta\Omega$ (in deg$^2$, right). {\it Right  panel}: eLISA capabilities of selecting among different MBH formation routes as a function of observation time. Plotted is the fraction of realizations in which one of the four investigated models (SE) is chosen over each of the other three models [LE (solid green), LC (long-dashed blue) and SC (short-dashed red)] at 95\% confidence level, as a function of observation time.}  
\label{fig7}
\end{figure*}
\begin{table*}[htb]
\begin{center}
\begin{tabular}{cc}
\begin{tabular}{|l|cccc|}
\hline
\hline
\multicolumn{5}{c}{Without spins}\\ \hline
 & SE & SC & LE & LC\\
\hline
SE & $\times$ &0.48 &0.99 &0.99\\
SC & 0.53& $\times$ &1.00 &1.00\\
LE & 0.01& 0.01& $\times$ &0.79\\
LC & 0.02& 0.02& 0.22& $\times$\\
\hline
\hline
\end{tabular}&
\begin{tabular}{|l|cccc|}
\hline
\hline
\multicolumn{5}{c}{With spins}\\ \hline
 & SE & SC & LE & LC\\
\hline
SE & $\times$ &0.96 &0.99 &0.99\\
SC & 0.13& $\times$ &1.00 &1.00\\
LE & 0.01& 0.01& $\times$ &0.97\\
LC & 0.02& 0.02& 0.06& $\times$\\
\hline
\hline
\end{tabular}
\end{tabular}
\end{center}
\caption{\label{tab1}Summary of all possible comparisons of the MBH evolution models. Results are for one year of observation with eLISA. We take a fixed confidence  level of $p=0.95$. The numbers in the upper-right half of each table show the fraction of realizations in which the {\it row} model is chosen at more than this confidence level when the {\it row} model is true. The numbers in the lower-left half of each table show the fraction of realizations in which the {\it row} model {\it cannot be ruled out} at that confidence level when the {\it column} model is true. In the left table we consider the trivariate $M$, $q$, and $z$ distribution of observed events; in the right table we also include the observed distribution of remnant spins, $a_r$.}
\end{table*}

While measurements of individual systems are extremely interesting and very useful for making, e.g., strong-field tests of GR, it is the properties of the whole set of MBHB mergers that are observed which will carry the most information for astrophysics. GW observations of multiple MBHBs may be used together to learn about their formation and evolution through cosmic history, as demonstrated by \cite{plowman11,sesana11}. We briefly provide here an illustrative example from \cite{amaro13a}. As argued above, in the general picture of MBH cosmic evolution, the population is shaped by the \emph{seeding process} and the \emph{accretion history}. \cite{amaro13a} therefore consider a set of 4 models with distinctive properties: (i) small seeds and extended (coherent) accretion (SE), (ii) light seeds and chaotic accretion (SC); (iii) large seeds and extended accretion (LE), (iv) large seeds and chaotic accretion (LC). Each model predicts a \emph{theoretical} distribution of coalescing MBHBs.  A given dataset $D$ of observed events can be compared to a given model $A$ by computing the likelihood $p(D|A)$ that the observed dataset $D$ is a realization of model $A$. When testing a dataset $D$ against a pair of models $A$ and $B$, one assigns probability $p_A=p(D|A)/(p(D|A)+p(D|B))$ to model $A$, and probability $p_B=1-p_A$ to model $B$. The probabilities $p_A$ and $p_B$ are a measure of the relative confidence one has in model $A$ and $B$, given an observation $D$. Setting a confidence threshold of $0.95$ one can count what fraction of the 1000 realizations of model $A$ yield a confidence $p_A>0.95$ when compared to an alternative model $B$. Results are shown in the left-hand panel of table \ref{tab1} for all pairs of models, assuming one year observation and circular non-spinning waveforms (i.e., for an extremely conservative waveform model). The vast majority of the pair comparisons yield a $95\%$ confidence in the true model for almost all the realizations, with the exception of comparisons LE to LC and SE to SC, i.e., comparisons among models differing by accretion mode only. This is because the accretion mode (efficient versus chaotic) particularly affects the spin distribution of the coalescing systems, which is not considered in the circular non-spinning waveform model. It is sufficient to add a measurement of the remnant spin parameter $a_r$ to make those pairs easily distinguishable (Right-hand panel of table \ref{tab1}). The right panel of figure \ref{fig7} shows the evolution of the fraction of correctly identified models as a function of observation time (no spin information included). Small versus large seed scenarios (SE vs LE and SE vs LC) can be easily discriminated after only 1 year of observation. 

\subsection{The nano-Hz regime: science with pulsar timing arrays}
\label{sec5.2}
PTAs are sensitive at much lower frequencies ($10^{-9}-10^{-7}$Hz), where the expected signal is given by a superposition of a large number of massive ($M>10^8\msun$), relatively nearby ($z<1$) sources overlapping in frequency. As argued in Section 3, at such low frequencies the properties of the MBHBs are likely to be severely affected by their coupling with their stellar and gaseous environment. In particular binaries can be highly eccentric, which might suppress the low frequency portion of the spectrum, crucial to PTA detection \cite{sesana13b}. Here we consider circular GW driven binaries for simplicity. The overall expected characteristic strain $h_c$ of the GW signal can be written as \cite{sesana08}
\begin{equation}
h_c^2(f) =\int_0^{\infty} 
dz\int_0^{\infty}d{\cal M}\, \frac{d^3N}{dzd{\cal M} d{\rm ln}f}\,
h^2(f),
\label{hch2}
\end{equation}
where $d^3N/dzd{\cal M} d{\rm ln}f$, is the comoving number of binaries emitting in a given logarithmic frequency interval with chirp mass and redshift in the range $[{\cal M},{\cal M}+d{\cal M}]$ and $[z, z+dz]$, respectively; and $h(f)$ is the inclination--polarization averaged strain given by equation (\ref{haverage}). 
\begin{figure*}
\centering
\vspace*{-0.0cm}
\begin{tabular}{cc}
\includegraphics[width=2.9in,clip=true]{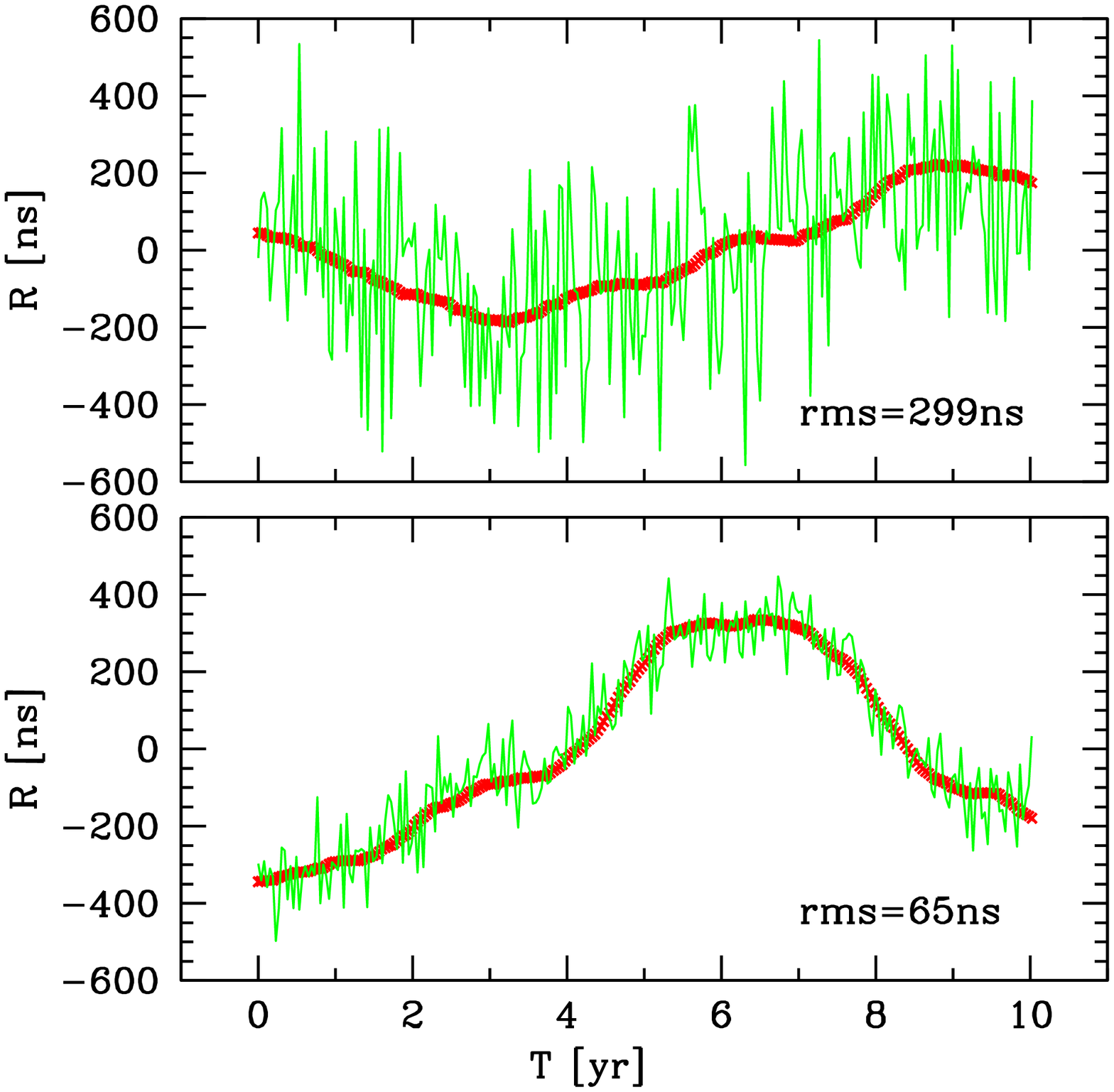}&
\includegraphics[width=2.9in,clip=true]{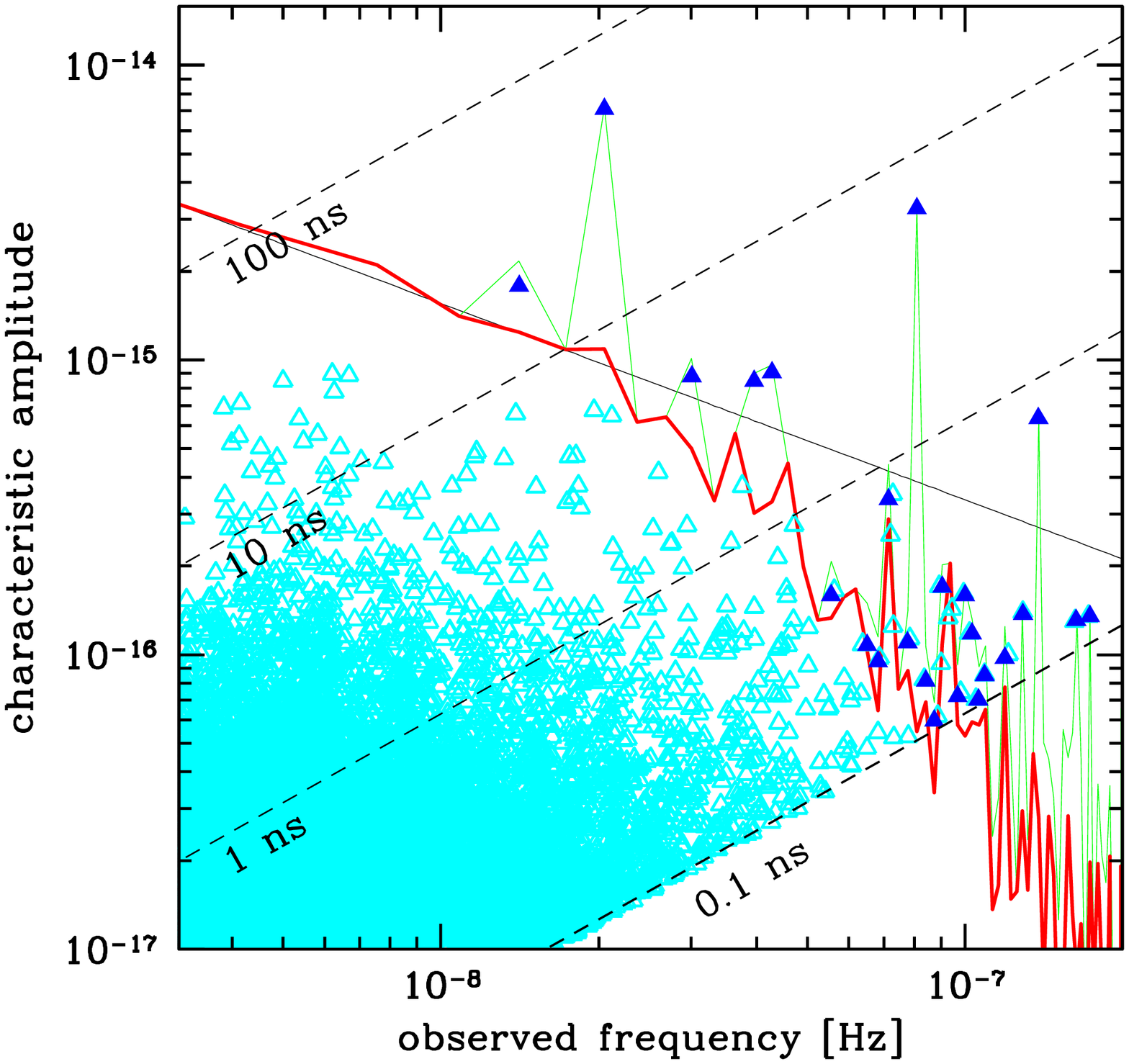}
\end{tabular}
\caption{{\it Left  panel}: simulated pulsar residuals $R$. The overall data, containing white noise with rms labeled in figure plus a Montecarlo realization of the GW signal, is represented in green; the GW signal only is given by the underlying thick red line. {\it Right panel}: Montecarlo realization of the GW signal expected in the PTA band; characteristic amplitude vs. frequency. Each cyan point represents an individual binary, and the overall signal is given by the green line. Blue triangles are potentially resolvable sources, and the red line is the level of the signal once the latter are subtracted: the unresolved background. The black solid line is the expected analytical $f^{-2/3}$ power law, and the black dashed lines represent different timing residual levels.}  
\label{fig8}
\end{figure*}
\begin{figure*}
\centering
\vspace*{-0.0cm}
\begin{tabular}{cc}
\includegraphics[width=2.9in,clip=true]{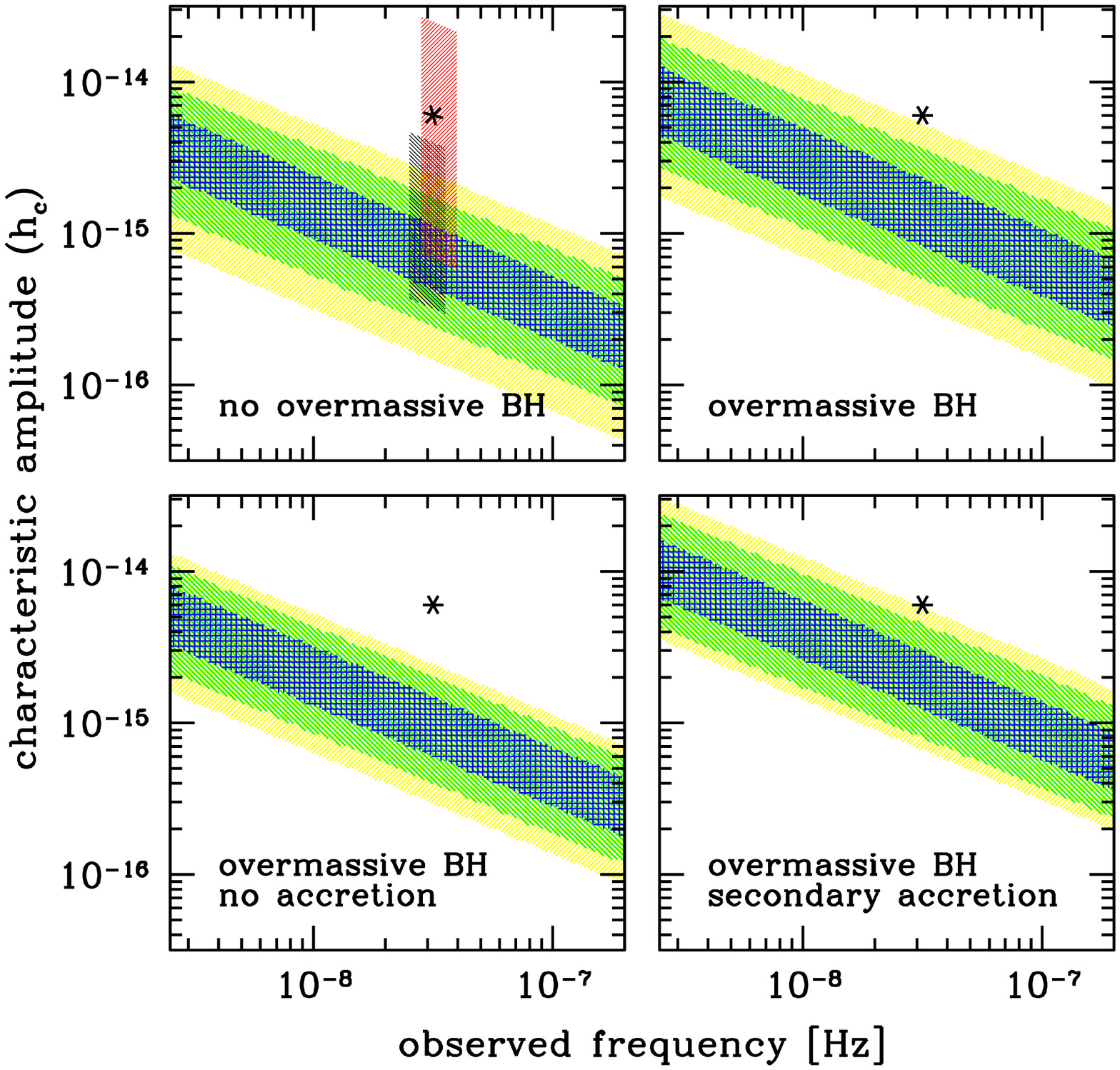}&
\includegraphics[width=2.9in,clip=true]{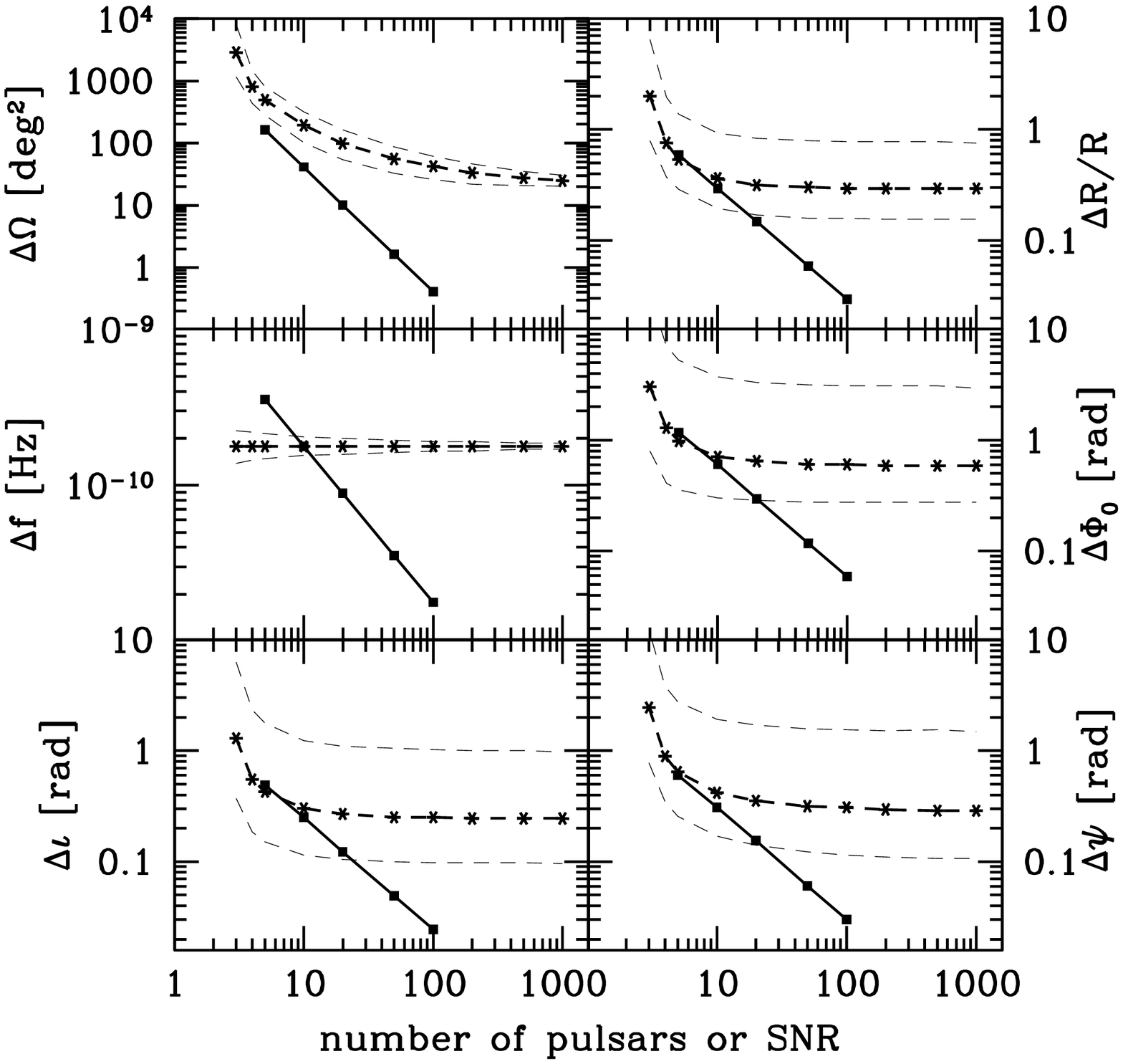}
\end{tabular}
\caption{{\it Left panel} Characteristic amplitude of the GW signal. Shaded areas represent the  $68\%$, $95\%$ and $99.7\%$ confidence levels given by  \protect\cite{sesana12}. In each panel, the black asterisk marks the best current limit from \protect\cite{vanh11}. Shaded areas in the upper left panel refer to the $95\%$ confidence level given by \protect\cite{mcwilliams12} (red) and the uncertainty range estimated by \protect\cite{sesana08}. 
{\it Right  panel}: Median expected statistical error on the source parameters. Each point (asterisk or square) is obtained by averaging over a large Monte Carlo sample of MBHBs.
In each panel, solid lines (squares) represent the median statistical error as a function of the total coherent SNR, assuming 100 randomly distributed pulsars in the sky; the thick dashed lines (asterisks) represent the median statistical error as a function of the number of pulsars for a fixed total SNR$=10$. In this latter case, thin dashed lines label the 25$^{\rm th}$ and the 75$^{\rm th}$ percentile of the error distributions (from \protect\cite{sesanavecchio10}).}  
\label{fig9}
\end{figure*}

The GW spectrum has a characteristic shape $h_c=A(f/1{\rm yr}^{-1})^{-2/3}$, where $A$ is the signal normalization at $f=1{\rm yr}^{-1}$, which depends on the details of the MBH binary population only. A Montecarlo realization of the signal is shown in figure \ref{fig8} for a selected MBHB population model. In the timing residual of each individual pulsar, the signal appears as a structured red noise (left panel), but a representation of the characteristic strain in the Fourier domain reveals the complexity of its nature (right panel). Although several millions of sources contribute to it, the bulk of the strain comes from few hundred sources only. Therefore, the signal is far from being a Gaussian isotropic background \cite{ravi12}; a handful of sources dominates the strain budget, and some of them might be individually identified. Detection techniques have been developed for stochastic signals \cite{hellings83,jen05,vanh09,anholm09} and individual sources \cite{ellis12,babak12,petiteau13}, and more sophisticated schemes accounting for signal anisotropy have recently been proposed \cite{cornish13,mingarelli13,taylor13}. In terms of level of the stochastic signal, recent works \cite{ravi12,mcwilliams12,sesana12} set a plausible range $3\times10^{-16}<A<4\times10^{-15}$, the upper limit being already in tension with current PTA measurement \cite{vanh11,demorest13}. This is shown in the left panel of figure \ref{fig9}, where observations are compared to theoretically predictions. Here, the difference between the top-left and the top-right panel is given by the recent upgrades in the MBH mass-host relation \cite{mcconnell13,grahamscott13} to include the overmassive black holes measured in brightest cluster galaxies (BCGs) \cite{fabian12}, that boosts the range of expected signal by a factor of two. In the lower panels instead, we consider two subset of the models featuring these upgraded relations: (i) those in which accretion does not occur prior to binary coalescence, and (ii) those in which accretion precedes the formation of the binary, and is more prominent on the secondary MBH \cite{callegari11}. In the latter case, binaries observed by PTA are much more massive (and with a larger $q$) implying a much larger (by almost a factor of three) signal.

An handful of sources might be bright enough to be individually resolved, and in this case some of their parameters can be determined according to the scheme described in Section \ref{sec4}. A pioneering investigation was performed by \cite{sesanavecchio10} assuming circular, non spinning monochromatic systems. In this case the waveform is function of 7 parameters only: the amplitude $R${\footnote{For monochromatic signals the two masses and the luminosity distance degenerate into a single amplitude parameter.}}, sky location $\theta,\phi$, polarization $\psi$, inclination $\iota$, frequency $f$ and phase $\Phi_0$, defining the parameter vector $\vec{\lambda} = \{R,\theta,\phi,\psi,\iota,f,\Phi_0\}$. Results about typical parameter estimation accuracy are shown in the right panel of figure \ref{fig9}. For SNR$=10$ The source amplitude is determined to a $20\%$ accuracy, whereas $\phi,\psi,\iota$ are only determined within a fraction of a radian. For bright enough sources (SNR$\approx10$) sky location within few tens to few deg$^2$ is possible (see also \cite{ellis12,petiteau13}), and even sub deg$^2$ determination, under some specific conditions \cite{lee11}. Even though this is a large chunk of the sky, these systems are extremely massive and at relatively low redshift ($z<0.5$), making any putative electromagnetic signature of their presence (e.g., emission periodicity related to the binary orbital period, peculiar emission spectra, peculiar K$\alpha$ line profiles, etc.) detectable \cite{sesana12,tanaka11}.  

\section{Conclusions}
\label{sec6}
We provided a general overview of massive black hole binaries as gravitational wave sources. MBHs are today ubiquitous in massive galaxies, and power luminous quasars up to $z>7$. Although they are believed to play a central role in the process of structure formation, their origin and early growth is largely unknown. According to our current understanding, MBHBs must form in large numbers along the cosmic history, providing the loudest sources of GWs in the Universe in a wide range of frequencies spanning from the sub-nano-Hz up to the milli-Hz. GWs carry precise information about the parameters of the emitting systems. We showed how those parameters are imprinted in the phase (and amplitude) modulation of the wave, and can therefore be efficiently extracted and determined to high accuracy with ongoing and future GW probes. From those we will learn about MBH formation and evolution through cosmic history, about the nature of the first BH seeds, their subsequent accretion history, and, more generally, about the early hierarchical structure formation at high redshift. We will also learn about the complex interplay of physical processes, including stellar and gas dynamics and GW emission, that leads to the dynamical formation and evolution of MBHBs. Direct GW detection will open a new era in MBH and MBHB astrophysics. 

\section*{Acknowledgments}
I would like to thank S. Babak, M. Dotti, F. Ohme and A. Petiteau for providing useful material. This work is supported by the DFG grant SFB/TR 7 Gravitational Wave Astronomy and by DLR (Deutsches Zentrum fur Luft- und Raumfahrt).

\section*{References}
\bibliographystyle{unsrt}
\bibliography{references}

\end{document}